\documentstyle[12pt]{article}
\def\PRL #1 #2 #3 {Phys.~Rev.~Lett.~{\bf #1}, #2 (#3)}
\def\PRD #1 #2 #3 {Phys.~Rev.~D~{\bf #1}, #2 (#3)}
\def\PLB #1 #2 #3 {Phys.~Lett.~{\bf B#1}, #2 (#3)}
\def\NPB #1 #2 #3 {Nucl.~Phys.~{\bf B#1}, #2 (#3)}

\newcommand{\gtap}{{\raise.3ex\hbox{$>$\kern-.75em\lower1ex\hbox{$\sim$}}}}
\newcommand{\ltap}{{\raise.3ex\hbox{$<$\kern-.75em\lower1ex\hbox{$\sim$}}}}

\begin{document}
\include{psfig}
\begin{titlepage}

\rightline{hep-ph/9707525} 
\medskip
\rightline{July 1997}
\bigskip\bigskip
\begin{center} {\Large \bf Nonperturbative Effects in Quarkonia Associated\\
\medskip
with Large Orders in Perturbation Theory} \\
\medskip
\bigskip\bigskip\bigskip\bigskip
{\large{\bf Martin C.~Smith, T.~Stelzer} and {\bf S.~Willenbrock}} \\ 
\medskip 
Department of Physics \\
University of Illinois \\ 1110 West Green Street \\  Urbana, IL\ \ 61801 \\
\bigskip 
\end{center} 
\bigskip\bigskip\bigskip

\begin{abstract}
We show that the perturbation series for quarkonia energies diverges
at large orders.  This results in a perturbative ambiguity in the
energy that scales as $e^{-1/a\Lambda}$ where $a$ is the Bohr radius
of quarkonium and $\Lambda$ is the QCD scale parameter.  This
ambiguity is associated with a nonperturbative contribution to the
energy from distances of order $\Lambda^{-1}$ and greater.  This
contribution is separate from that of the gluon condensate.
\end{abstract}

\end{titlepage}

\newpage

The states of quarkonia which have a size much less than the
characteristic scale of nonperturbative QCD, $\Lambda^{-1}$, can be
accurately described by QCD perturbation theory, just as positronium
is accurately described by QED perturbation theory \cite{CL}.  The
quarkonium wave function is nonzero at long distances, of order
$\Lambda^{-1}$ and greater, so the states of quarkonia are also
influenced by nonperturbative physics.  In this article we show that
perturbation theory is aware of this nonperturbative effect via its
large-order behavior.

Nonperturbative effects in quarkonia have previously been
paramaterized in terms of vacuum condensates.  At short distances, the
leading such effect is due to the gluon condensate \cite{V,L}.  We
argue that this effect is separate from the nonperturbative effect we
investigate, which is due to the nonzero wave function at long
distances.

That perturbation theory is aware of (some) nonperturbative effects
via its large-order behaviour is familiar from the study of infrared
renormalons \cite{tH,M}. Renormalons correspond to factorial growth of
the coefficients of the perturbation series, and are associated with
nonperturbative contributions of order $(\Lambda/Q)^p$, where $Q$ is a
characteristic momentum and $p$ is a positive integer.  Our analysis
is in the spirit of infrared renormalons.  We, however,find
perturbative coefficients that grow less rapidly than factorially, and
which are associated with a nonperturbative contribution of order
$e^{-Q/\Lambda}$, rather than a power of $(\Lambda/Q)$.

At short distances, quarkonia can be described by the QCD analogue of
the Coulomb potential \cite{CL},
\begin{equation}
V(r) = -\frac{4}{3}\frac{\alpha_s(1/r)}{r}
\label{pot}
\end{equation}
with a coupling that depends on the distance between the quark and the
antiquark.  At distances of order $\Lambda^{-1}$ and greater the
coupling is large, and Eq.~(\ref{pot}) is invalid.  To study the
effects of the running coupling perturbatively, we expand the
evolution equation for the coupling as an infinite sum of terms which
depend on the coupling at an arbitrary fixed scale~$\mu$, $\alpha_s
\equiv \alpha_s(\mu)$.
\begin{eqnarray}
\alpha_s(1/r) & = & \frac{\alpha_s}{1-\beta\alpha_s \ln \mu r}
\nonumber \\ & = & \alpha_s(\mu) \sum_{k=0}^\infty (\beta \alpha_s \ln
\mu r)^k \;.
\label{run}
\end{eqnarray}
where
\begin{equation}
\beta \equiv \frac{1}{2\pi}\left(11-\frac{2}{3}N_f\right)
\end{equation}
is the one-loop beta function for $N_f$ flavors of light quarks.
Inserting Eq.~(\ref{run}) into Eq.~(\ref{pot}) gives
\begin{equation}
V(r) = -\frac{4}{3}\frac{\alpha_s}{r} + \sum_{k=1}^\infty V^k(r)
\label{potpert}
\end{equation}
with
\begin{equation}
V^k(r) \equiv -\frac{4}{3}\frac{\alpha_s}{r}(\beta\alpha_s\ln \mu r)^k
\end{equation}
Diagrammatically, the potential in Eq.~(\ref{potpert}) corresponds to
the sum of vacuum-polarization insertions in the tree-level potential,
as shown in Fig.~1.  In this sense, our analysis is similar to that of
infrared renormalons.

\begin{figure}
\centerline{\psfig{figure=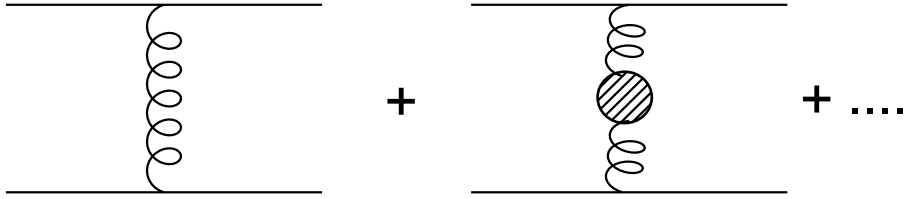,width=5in}}
\caption{Quarkonium formation via the exchange of a gluon.  The
Coulomb potential is modified by the insertion of vacuum-polarization
subdiagrams in the gluon propagator.}
\end{figure}

The first term in Eq.~(\ref{potpert}) is the QCD Coulomb potential
with a fixed coupling.  We take this as the unperturbed potential. Its
solution involves the usual Coulomb energies and wave functions.  We
regard the remaining terms as perturbing potentials of order
$(\beta\alpha_s)^k$, and evaluate their contribution using ordinary
nonrelativistic time-independent perturbation theory.

To evaluate the contribution of the $k^{\rm th}$ perturbing potential
to the energy of the state with principal quantum number $n$, orbital
quantum number $l$, and azimuthal quantum number $m$,\footnote{Since
the potential is radially symmetric, the states with different
azimuthal quantum numbers are degenerate.} one evaluates the matrix
element
\begin{equation}
\Delta E_{nlm}^k = <\psi_{nlm}|V^k|\psi_{nlm}> \;.
\label{delta}
\end{equation}
For simplicity we momentarily restrict our attention to the ground
state. For quarks of mass $m_Q$, the ground-state wavefunction and
energy are
\begin{equation}
\psi_{100}(r) = \frac{1}{\sqrt{\pi a^3}}e^{-r/a}
\label{wave}
\end{equation}
\begin{equation}
E = -\frac{m_Q}{4} \left( \frac{4}{3} \alpha_s \right)^2 =
-\frac{2\alpha_s}{3a}
\end{equation}
where $a$ is the Bohr radius,
\begin{equation}
a = \left(\frac{m_Q}{2}\frac{4}{3}\alpha_s\right)^{-1} \;.
\label{bohr}
\end{equation}
(The calculation for arbitrary $n,l,m$ is neither more difficult nor
more enlightening; we present the result later.)  Inserting
Eq.~(\ref{wave}) into Eq.~(\ref{delta}), dividing by $E$, and
integrating over angles yields
\begin{equation}
\frac{\Delta E^k}{E} = \frac{8}{a^2} (\beta\alpha_s)^k \int_0^\infty
dr\, r\, \ln^k \mu r\, e^{-2r/a}
\label{ekr}
\end{equation}
Making the change of variables $z=\mu r$ and setting $c = \mu a/2$
gives
\begin{equation}
\frac{\Delta E^k}{E} = \frac{2}{c^2} (\beta\alpha_s)^k I^k
\label{ek}
\end{equation}
where
\begin{eqnarray}
I^k \equiv \int_0^\infty dz\, z\, \ln^k z\, e^{-z/c} = \int_0^\infty
dz\, z\, e^{k \ln \ln z - z/c}
\label{int}
\end{eqnarray}

For large $k$, the integral of Eq.~(\ref{int}) is dominated by the
region where the argument of the exponential is a maximum.  We denote
this maximum by $z_0$, defined by the implicit equation
\begin{equation}
z_0 \ln z_0 = kc \;.
\label{z0}
\end{equation}
We approximate the integral by the saddle-point method, which yields
\begin{equation}
I^k \approx \sqrt{2\pi c}\, z_0^{3/2} \ln^k z_0\, e^{-z_0/c} \;.
\label{ik}
\end{equation}
Thus Eq.~(\ref{ek}) becomes
\begin{equation}
\frac{\Delta E^k}{E} \approx \sqrt{8\pi}\, \left( \frac{z_0}{c}
\right)^{3/2} (\beta\alpha_s \ln z_0)^k\, e^{-z_0/c} \;.
\label{ekz0}
\end{equation}

The series for the energy correction, whose $k^{\rm th}$ term is given
by Eq.~(\ref{ekz0}), is divergent.  For large $k$, the size of the
$k^{\rm th}$ term is governed by the factor
\begin{equation}
S^k = (\beta\alpha_s \ln z_0)^k\, e^{-z_0/c} \;.
\label{sk}
\end{equation}
The term where the series begins to diverge is obtained by finding the
minimum of this function with respect to $k$ (keeping in mind that
$z_0$ implicitly depends on $k$ via Eq.~(\ref{z0})):
\begin{equation}
0 = \frac{d\ln S^k}{dk} = \ln (\beta\alpha_s) + \ln \ln z_0 +
\frac{k}{z_0 \ln z_0} \frac{dz_0}{dk} - \frac{1}{c}\frac{dz_0}{dk}
\end{equation}
Using Eq.~(\ref{z0}), we find that the last two terms cancel.  Solving for 
$z_0$ gives 
\begin{equation}
z_0 = e^{1/\beta\alpha_s} = \frac{\mu}{\Lambda}
\label{z0beta}
\end{equation}
where the last relation follows from the definition of the QCD scale
parameter $\Lambda$:
\begin{equation}
\alpha_s \equiv \alpha_s(\mu) = \left( \beta \ln \frac{\mu}{\Lambda}
\right)^{-1} \;.
\end{equation}
Recalling that $z=\mu r$, Eq.~(\ref{z0beta}) tells us that the series
begins to diverge when the integral in Eq.~(\ref{ekr}) is dominated by
distances of order $\Lambda^{-1}$.  This feature is reminiscent of
infrared renormalons.

Within the context of perturbation theory, the best one can do is to
truncate the series at the minimum term.  We take the size of this
term as an estimate of the inherent ambiguity in the perturbative
calculation of the energy.  Inserting Eq.~(\ref{z0beta}) into
Eq.~(\ref{ekz0}) yields
\begin{eqnarray}
\frac{\delta E}{E} & \approx & \sqrt{8\pi}\, \left( \frac{z_0}{c}
  \right)^{3/2} e^{-z_0/c} \nonumber \\ & = & \sqrt{8\pi}\, \left(
  \frac{2}{a\Lambda} \right)^{3/2} e^{-2/a\Lambda} \;.
\end{eqnarray}
The analogous calculation for the state with quantum numbers $n,l,m$
with wavefunction\footnote{We use the notation and conventions of
Ref.~\cite{laguerre}.  The $L_{n-l-1}^{2l+1}$ are associated Laguerre
polynomials.}
\begin{equation}
\psi_{nlm}(r,\theta,\phi) = \frac{2}{n^2 a^{3/2}}
\left(\frac{(n-l-1)!}{(n+l)!}\right)^{1/2}
\left(\frac{2r}{na}\right)^l L_{n-l-1}^{2l+1}(2r/na) e^{-r/na}
Y_l^m(\theta,\phi)
\end{equation}
and energy
\begin{equation}
E_n = -\frac{m_Q}{4n^2} \left( \frac{4}{3} \alpha_s \right)^2 =
-\frac{2\alpha_s}{3n^2a}
\end{equation}
yields the result
\begin{equation}
\frac{\delta E_{nl}}{E_n} \approx \sqrt{8 \pi}\,
\frac{(n-l-1)!}{(n+l)!}  \left( \frac{2}{na\Lambda}
\right)^{2l+\frac{3}{2}} [L_{n-l-1}^{2l+1}(2/na\Lambda)]^2\,
e^{-2/na\Lambda}
\label{dele}
\end{equation}
(with $a$ given by Eq.~(\ref{bohr})).  This is the main result of the
paper.

In a full QCD calculation, the ambiguity in the perturbative
calculation of the energy, Eq.~(\ref{dele}), would be cancelled by a
related ambiguity in the nonperturbative contribution to the energy,
resulting in an unambiguous expression.  This cancellation between
perturbative and nonperturbative ambiguities is in the same spirit as
the analogous cancellation which occurs for infrared renormalons
\cite{M}.  Thus we learn that quarkonia energies have a
nonperturbative contribution which scales with $na\Lambda$ like
Eq.~(\ref{dele}).  However, it is impossible to calculate the
coefficient of this nonperturbative contribution within the context of
perturbation theory.

The ambiguity in the perturbative calculation of the energy,
Eq.~(\ref{dele}), is nearly proportional to the probability density
that the quarks in quarkonia are separated by a distance of
$\Lambda^{-1}$, which is given by $\rho^2|\psi^{nlm}(\rho)|^2 \sim
\rho^{2l+2} [L_{n-l-1}^{2l+1}(\rho)]^2 e^{-\rho}$, with $\rho =
2/na\Lambda$.  The average separation for the quarks is roughly $n^2
a$.  The ambiguity increases as $\rho$ decreases, {\it i.e.}, as the
size of the state increases, until $\rho \sim O(n)$, at which point it
apparently begins to decrease.  However, this value of $\rho$
corresponds to $n^2 a \approx \Lambda^{-1}$, where our perturbative
approach is unreliable, and Eq.~(\ref{dele}) should not be trusted.

The ambiguity in the perturbative calculation of the energy,
Eq.~(\ref{dele}), is proportional to $e^{-2/na\Lambda}$ rather than a
power of $a\Lambda$ (as would arise from an infrared renormalon)
because the series for the energy correction, Eq.~(\ref{ekz0}),
diverges less rapidly than factorially.  An approximate solution to
Eq.~(\ref{z0}) is $z_0 \approx kc/\ln kc$; thus the factor $\ln^k
z_0\, e^{-z_0/c}$ in Eq.~(\ref{ekz0}) is approximately $\ln^k k\,
e^{-k}$, which grows less rapidly than factorially, $k! \sim k^k\,
e^{-k}$.  A similarly divergent series, resulting in a perturbative
ambiguity proportional to $e^{-Q(1-\tau)/\Lambda}$, appears in the
calculation of soft-gluon resummation in the production of an object
of mass $Q$ from a hadron collision of total energy $\sqrt S$
($\tau=Q^2/S$) \cite{CMNT}.

A nonperturbative contribution to quarkonia energies from the gluon
condensate has been previously studied \cite{V,L}.  This contribution
is associated with an infrared renormalon arising from emission and
absorption of soft gluons, as shown in Fig.~2. The perturbative
ambiguity from this renormalon is cancelled by a related ambiguity in
the gluon condensate \cite{M}.  The gluon-condensate contribution to
the quarkonia energies is \cite{V,L}
\begin{equation}
\Delta E_{nl} = \frac{\pi}{16} m_Q n^6 a^4 <\alpha_s G^2>
\epsilon_{nl}
\label{delcond}
\end{equation}
where $\epsilon_{nl}$ are numbers of order unity.  Since $<\alpha_s
G^2>$ is of order $\Lambda^4$, this contribution is proportional to
$(a\Lambda)^4$.  Both the physical origin of Eq.~(\ref{delcond}) and
its functional dependence on $a\Lambda$ show that it is a completely
separate nonperturbative effect from the one considered in this paper.

\begin{figure}
\centerline{\psfig{figure=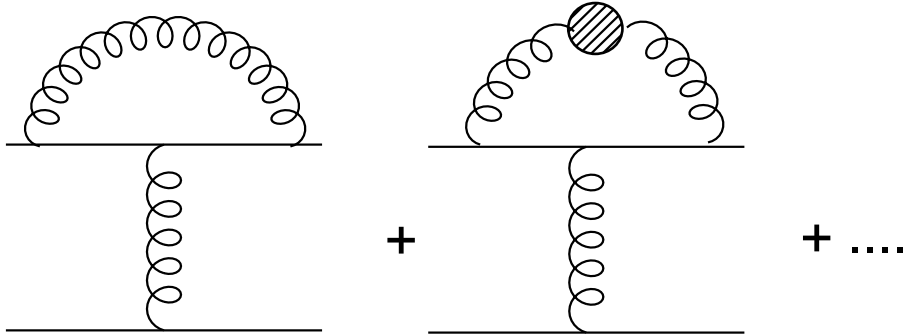,width=5in}}
\caption{A higher-order contribution to the quarkonium energy from the
emission and absorption of gluons.  The insertion of
vacuum-polarization subdiagrams in the gluon propagator leads to an
infrared renormalon associated with the gluon condensate.}
\end{figure}

The gluon condensate has been considered as the leading
nonperturbative contribution to quarkonia in various analyses
\cite{TY,AY,STY,V2,PS}.  For states which are small compared with
$\Lambda^{-1}$, the gluon condensate, proportional to $(a\Lambda)^4$,
is formally more important than the nonperturbative effect discussed
in this paper, proportional to $e^{-2/na\Lambda}$.  However, as $n^2
a$ approaches $\Lambda^{-1}$, the relative size of the
gluon-condensate contribution and the nonperturbative contribution
considered in this paper are unknown, since we do not know the
normalization of the latter.  It is therefore important to establish
the size of this nonperturbative effect.

In this paper we have shown that the calculation of quarkonia energies
is divergent at large orders in perturbation theory.  The associated
ambiguity in the perturbative result, Eq.~(\ref{dele}), is
proportional to $e^{-2/na\Lambda}$, where $a$ is the Bohr radius, $n$
is the principal quantum number, and $\Lambda$ is the QCD scale
parameter.  This ambiguity is associated with a nonperturbative
contribution to the energy, coming from distances of order
$\Lambda^{-1}$ and greater, which scales with $na\Lambda$ like
Eq.~(\ref{dele}).  This is a separate nonperturbative effect from the
gluon-condensate contribution to quarkonia energies.  It is important
to establish the size of this nonperturbative effect to judge its
impact on quarkonia energies.

\section*{Acknowledgements}

\indent\indent We are grateful for conversations with R.~Akhoury,
E.~Braaten, A.~El-Khadra, A.~Hoang, A.~Kronfeld, R.~Leigh,
P.~Mackenzie, M.~Mangano, M.~Olsson, J.~Rosner, and J.~Stack.  This
work was supported in part by Department of Energy grant
DE-FG02-91ER40677.  S.~W.~thanks the Aspen Center for Physics where
part of this work was performed.  We gratefully acknowledge the
support of a GAANN fellowship, under grant number DE-P200A10532, from
the U.~S.~Department of Education for M.~S.


\begin{thebibliography}{99}

\bibitem{CL} W.~Caswell and G.~P.~Lepage, \PLB 167 437 1986 .

\bibitem{V} M.~Voloshin, \NPB 154 365 1979 ; Yad.~Fiz.~{\bf 36}, 247 (1982) 
[Sov.~J.~Nucl.~Phys.~{\bf 36}, 143 (1982)].

\bibitem{L} H.~Leutwyler, \PLB 98 447 1981 .

\bibitem{tH} G.~'t Hooft, in {\sl The Whys of Subnuclear Physics},
Proceedings of the International School of Subnuclear Physics, Erice,
1977, ed.~A.~Zichichi (Plenum, New York, 1979), p.~943; A.~Mueller, in
{\sl QCD - 20 Years Later}, Proceedings of the Workshop, Aachen,
Germany, 1992, eds.~P.~Zerwas and H.~Kastrup (World Scientific,
Singapore, 1993), Vol.~1, p.~162.

\bibitem{M} A.~Mueller, in {\sl QCD 20 Years Later}, Aachen, June 9-13, 1992,
eds.~P.~Zerwas and H.~Kastrup (World Scientific, Singapore, 1993), p.~162.

\bibitem{laguerre} J.~Powell and B.~Crasemann, {\sl Quantum Mechanics}
(Addison-Wesley, Menlo Park, 1961); 
I.~Gradshteyn and I.~Ryzhik, {\sl Table of Integrals, Series, and Products} 
(Academic Press, San Diego, 1965).
 
\bibitem{CMNT} S.~Catani, M.~Mangano, P.~Nason, and L.~Trentadue,
\NPB 478 273 1996 .

\bibitem{TY} S.~Titard and F.~Yndur\'{a}in, \PRD 49 6007 1994 ; \PRD
51 6348 1995 ; \PLB 351 541 1995 .

\bibitem{AY} K.~Adel and F.~Yndur\'{a}in, \PRD 52 6577 1995 .

\bibitem{STY} Yu.~Simonov, S.~Titard, and F.~Yndur\'{a}in, \PLB 354
435 1995 .

\bibitem{V2} M.~Voloshin, Int.~J.~Mod.~Phys.~{\bf A10}, 2865, 1995.

\bibitem{N} S.~Narison, \PLB 387 162 1996 .

\bibitem{PS} A.~Pineda and J.~Soto, \PRD 53 3983 1996 ; \PRD 54 4609 1996 .

\end{thebibliography}
\end{document}